\documentstyle[preprint,aps]{revtex}
\def\lan{\langle}
\def\ran{\rangle}
\begin{document}
\draft
\title{\bf{\LARGE{Stochastic resonance and nonlinear response in a
dissipative quantum two-state system }}}
\author{T. P. Pareek, Mangal C. Mahato and A. M. Jayannavar  }
\address{Institute of Physics, Sachivalaya Marg, Bhubaneswar - 751 005, India}
\maketitle
\vspace{0.1in}
\begin{abstract}

We study the dynamics of a dissipative two-level, system driven
by a monochromatic ac field, starting from the usual spin-boson
Hamiltonian. The quantum Langevin equations for the spin variables
are obtained. The amplitude of the coherent oscillations in the
average position of the particle is studied in the high
temperature limit. The system exhibits quantum stochastic
resonance in qualitative agreement with earlier numerical results.

\end{abstract}
\vspace{0.4in}
\pacs{PACS numbers : 73.40.Gk,05.30.-d,05.40.+j,03.65.-w.}
%\narrowtext
%\thispagestyle{empty}
\newpage
\section{Introduction}

Stochastic resonance is a nonlinear phenomena. It has been
predicted and experimentally observed in damped classical
systems[1-4]. It is characterized by a maximum in the response of the
system to an external input signal as a function of noise
strength at the input signal frequency. There have been some
recent studies in stochastic resonance in damped quantum double-well
systems as well [5-9]. Recently, Makarov and Makri have studied the
phenomena in a dissipative two-level system numerically using an
iterative path integral scheme and tried to understand the
results analytically[8,9]. They demonstrate that it is possible to
induce and maintain large amplitude coherent oscillations by
exploiting the phenomena of stochastic resonance in  quantum
systems. In their treatment noise strength is varied, through the
coupling to heat bath, to obtain stochastic resonance. The
maximum in response is also obtained with respect to the driving
field strength indicating breakdown of linear response theory.
However, their analytical treatment involves several
approximations. They start with the evolution of dynamical
variables of a system using Heisenberg equation of motion for a
given spin-boson (system-bath) Hamiltonian in the presence of a
monochromatic field. In their weak coupling approximation the bath
causes stochastic energy
fluctuations which are unaffected by the dynamics of the
two-level system . To obtain the steady state average
position of the particle $\langle \sigma_{z} \rangle$ they
require steady state solution for the population difference
between the two-level-system eigenstates $\langle \sigma_{x}
\rangle$ . The required expression for $\langle \sigma_{x}
\rangle$ is then taken from the solutions of the phenomenological
optical Bloch equations in the rotating wave approximation. The
expression for $\langle \sigma_{z} \rangle$ thus obtained with these
approximations seems to be in good agreement with their
numerical simulations.

\par  In the present work we study stochastic resonance
analytically by systematically deriving the quantum Langevin
equation for the system by eliminating bath variables[10-13]. The
response of the system to an oscillating field is obtained in
the high temperature limit. This limit corresponds to treating
the random force operator as a classical c-number variable. Our
results are in qualitative agreement with earlier works.

\par In section II the derivation of quantum Langevin equation
is presented. The spin Bloch equations are obtained and
the stationary solution of the relevant variables is given
in sec. III. The last section IV is devoted to
results and discussion.

II. {\bf The Quantum Langevin equation for system variables}

\par To obtain the Quantum Langevin equations for system variables we consider 
a symmetric two-level system interacting linearly with a bath of
harmonic oscillators in the presence of a time dependent external
monochromatic  field ${V_0}cos(\omega t)$.
The total Hamiltonian is given by [8,9]
\begin{equation}
\label{h}
H= -\hbar\Delta_0 \sigma_x + \sum \hbar \omega_k a^\dagger_{k} a_{k}
+ \sum {s_0} g_k (a_k + a^\dagger_{k}) \sigma_z + V_0 \cos(\omega t)\sigma_z,
\end{equation}
\noindent where $\sigma$ 's are the Pauli matrices. $a_{k}$ and
$a^\dagger_{k}$  are annihilation and creation operators for
bath variables and $g_{k}$ is the coupling constant and $2s_0$ is
the distance between the two wells. The
population difference between the two eigenstates of the
two-level system, which are separated by $2\hbar\Delta_{0}$, is
given by $\langle \sigma_{x} \rangle$. The average value of
$\langle \sigma_{z} \rangle$ represents the average position of
the particle, $\langle \sigma_{z} \rangle$ being +1 or -1 is
equivalent to the particle in the right or the left well, respectively.

\par The quantum equation of motion for any dynamical variable
$A$ can be obtained from the following evolution equation
\begin{equation}
\label{cu}
\frac{dA}{dt} = \frac{i}{\hbar} [H , A],
\end{equation}
\noindent where [-- , --] indicates commutator. The commutaton
relations among the spin variables are given by
\begin{equation}
\label{sgm}
[\sigma_{x} , \sigma_{y}] = 2 i \sigma_{z},
\end{equation}
\noindent and their cyclic permutations. Moreover,
\begin{equation}
\label{aa}
[a_{k^\prime}, a^{\dagger}_{k}] = \delta_{{k}{k^\prime}}.
\end{equation}
\par Using equations (1)-(4) one can readily write down the equations
of motion for the variables as
\begin{eqnarray}
\label{eqm1}
\frac{d\sigma_x}{dt}&=& -2\sum\frac{s_{0}g_k}{\hbar}(a_{k}(t)+a^\dagger_{k}(t))
\sigma_{y}(t)-\frac{2V_{0}}{\hbar}cos(\omega t)\sigma_{y}(t), \; \;
\end{eqnarray}

\begin{equation}
\label{eqm2}
\;\;\;\;\;\;\;\;\; \frac{d\sigma_y}{dt}= 2\Delta_{0}\sigma_{z}(t)+
2\sum\frac{s_{0}g_k}{\hbar}(a_{k}(t)+a^\dagger_{k}(t))
\sigma_{x}(t)+\frac{2V_{0}}{\hbar}cos(\omega t)\sigma_{x}(t),
\end{equation}
\begin{equation}
\label{eqm3}
\frac{d\sigma_z}{dt}=-2\Delta_{0}\sigma_{y}(t), \;\;\;\;\;\;\;\;\;\;\;\;\;\;\;
\end{equation}

\begin{equation}
\label{eqm4}
\frac{da_{k}}{dt}=-i\omega_{k}a_{k}(t)-i\frac{s_{0}g_k}{\hbar}\sigma_{z}(t),
\end{equation}
\noindent and,
\begin{equation}
\label{eqm5}
\frac{da^\dagger_{k}}{dt}=+i\omega_{k}a_{k}(t)+i\frac{s_{0}g_k}{\hbar}\sigma_{z}(t).
\end{equation}
\noindent Equations (\ref{eqm4}) and (\ref{eqm5}) are linear and therefore can be
explicitly integrated, to obtain
\begin{equation}
\label{eqm41}
a_{k}(t)=a_{k}(0)e^{-i\omega_{k}t}
-i\frac{s_{0}g_k}{\hbar}\int_{0}^{t}\sigma_{z}(t^\prime)
e^{-i\omega_{k}(t-t^\prime)}dt^{\prime},
\end{equation}

\begin{equation}
\label{eqm51}
a^{\dagger}_{k}(t)=a^{\dagger}_{k}(0)e^{i\omega_{k}t}
+i\frac{s_{0}g_k}{\hbar}\int_{0}^{t}\sigma_{z}(t^\prime)
e^{i\omega_{k}(t-t^\prime)}dt^{\prime}.
\end{equation}
\par Where $a_{k}(0)$ and $a^{\dagger}_{k}(0)$ are the bath
operator values at the initial time $t=0$. Substituting
for $a_{k}(t)$ and $a^{\dagger}_{k}(t)$ from equations (\ref{eqm41})
 and (\ref{eqm51})
in equations (\ref{eqm1}) and (\ref{eqm2}),
we obtain the following Quantum Langevin equations
for the system variables,
\begin{eqnarray}
\label{sx2}
\frac{d\sigma_{x}}{dt}&=& -\frac{2s_{0}}{\hbar}{F(t)}\sigma_{y}(t)
-\frac{2V_{0}}{\hbar}\cos(\omega t)\sigma_{y}(t) \nonumber\\[0.5cm] 
&& +\frac{1}{2}\left[\sum_{k}\frac{4s_{0}^{2} g_{k}^2}{\hbar^2}
\left\{\int_{0}^{t}\sin(\omega_{k}(t-t^\prime)) 
\sigma_{z}(t^\prime)dt^{\prime}\right\}\sigma_{y}(t) \right. \nonumber\\[0.5cm] 
&& \left. + \sum_{k}\frac{4s_{0}^{2} g_{k}^2}{\hbar^2}\sigma_{y}(t) 
\left\{\int_{0}^{t}\sin(\omega_{k}(t-t^\prime)) 
\sigma_{z}(t^\prime)dt^{\prime}\right\}\right],
\end{eqnarray}
\begin{eqnarray}
\label{sy2}
\frac{d\sigma_{y}}{dt}&=&2\Delta_{0}\sigma_{z}(t)
+\frac{2s_{0}\sigma_{x}(t)}{\hbar}
{F(t)}+\frac{2V_0}{\hbar}cos(\omega t)\sigma_{x} \nonumber\\[0.5cm]
&& -\frac{1}{2}\left[\sum_{k}\frac{4 s_{0}^{2}g_{k}^2}{\hbar^2}
\sigma_{x}(t)\left\{\int_{0}^{t}
\sin(\omega_{k}(t-t^\prime)
\sigma_{z}(t^\prime)dt^{\prime}\right\} \right. \nonumber\\[0.5cm]
&& \left. -\frac{1}{2}\sum_{k}\frac{4s_{0}^{2}g_{k}^2}{\hbar{^2}}
\left\{\int_{0}^{t}
\sin(\omega_{k}(t-t^\prime)
\sigma_{z}(t^\prime)dt^{\prime}(t)\right\}\sigma_{x}(t)\right],
\end{eqnarray}
\noindent Here $F(t)$ is given by
\begin{equation}
\label{rand}
{F(t)}=\sum_{k}g_{k}\large(a_{k}(0)e^{-i\omega_{k}t}+
a^\dagger_{k}(0)e^{i\omega_{k}t}\large).
\end{equation}

\par As the dynamical operators
($a_{k}(0)$ , $a_{k}{^\dagger}(0)$) of the bath are distributed
in accordance with the statistical equilibrium distribution for
given temperature $T$, $F(t)$ is referred to as  Langevin
operator noise term [12,13]. The integrals in equations (\ref{sx2}) 
and (\ref{sy2}) can
be integrated by parts leading to
\begin{eqnarray}
\label{sx3}
\frac{d\sigma_{x}}{dt}&=& - \frac{2s_{0}}{\hbar}{F(t)}\sigma_{y}(t)
-\frac{2V_{0}}{\hbar}\cos(\omega t)\sigma_{y}(t) \nonumber\\[0.5cm]
&& +\frac{1}{2}\sum_{k}\frac{4s_{0}^{2}g_{k}^2}{\hbar{^2}\omega_k}
\int_{0}^{t}\left\{\frac{d\sigma_{z}(t^\prime)}{dt^\prime}\sigma_{y}(t)
+\sigma_{y}(t) {\frac{d\sigma_{z}(t^\prime)}{dt^\prime}} \right\}
cos(\omega_{k}(t-t^\prime)) dt^{\prime} \nonumber\\[0.5cm]
&& -\frac{1}{2}\sum\frac{4s_{0}^{2}g_{k}^2}{\hbar^{2}\omega_k}
\left\{\sigma_{z}(0)\sigma_{y}(t)
+\sigma_{y}(t) \sigma_{z}(0)\right\} cos(\omega_{k}t),
\end{eqnarray}
\begin{eqnarray}
\label{sy3}
\frac{d\sigma_{y}}{dt}&=&2\Delta_{0}\sigma_{z}(t) +
\frac{2s_{0}}{\hbar} {F(t)}\sigma_{x}(t)
+\frac{2V_0}{\hbar}cos(\omega t)\sigma_{x}(t) \nonumber\\[0.5cm]
&& -\frac{1}{2}\sum_{k}\frac{4s_{0}^{2}g_{k}^2}{\hbar{^2}\omega_k}
\int_{0}^{t}\left\{\frac{d\sigma_{z}(t^\prime)}{dt^\prime}
\sigma_{x}(t) +\sigma_{x}(t) \frac{d\sigma_{z}(t^\prime)}{dt^\prime} \right\}
cos(\omega_{k}(t-t^\prime)) dt^{\prime} \nonumber\\[0.5cm]
&&+\frac{1}{2}\sum\frac{4s_{0}^{2}g_{k}^2}{\hbar^{2}\omega_k}
\left\{\sigma_{z}(0)\sigma_{x}(t)
+\sigma_{x}(t) \sigma_{z}(0)\right\}cos(\omega_{k}t).
\end{eqnarray}
\par In equations (\ref{sx3}) and (\ref{sy3})
$G(t-t^{\prime})=\sum(4s_{0}^{2}g_{k}^2/\hbar^{2}\omega_k)cos(\omega_{k}(t-t^{\prime}))$
represents a damping or memory kernel [10]. The last two
terms in equation (\ref{sx3}) and (\ref{sy3}) are transient terms which can be
neglected if one assumes Ohmic spectral density for bath
oscillators [10,11],i.e.,
\begin{equation}
\label{rho}
\rho(\omega)=\frac{\pi}{2}\sum\frac{4s_{0}^{2}g_{k}^2}{\hbar^2}
\delta(\omega-\omega_k)=\alpha\omega exp(-\frac{\omega}{\omega_c}),
\end{equation}
\noindent where $\alpha$ is a dimension less dissipation coefficient 
(or Kondo parameter). Notice that the form of the spectral density
is not bounded from above and hence for physical reasons one
introduces an upper cut-off frequency  $\omega_c$, namely 
$\rho(\omega)=\alpha\omega{e^{-(\omega/\omega_c)}}$, such that the
frequency scale $\omega_c$ is assumed to be much larger than the
characteristic frequencies of the problem. With this form of
spectral density one can readily show that the transient terms do
survive up to a time scale $(1/\omega_c)$, which can be made
arbitrarily small and thus can be ignored [11]. 
Thus for the long time behaviour and for the Ohmic
spectral density the equations (\ref{sx3}) and (\ref{sy3}) get further simplified
and we arrive at the equations of motion for spin variables 
as
\begin{eqnarray}
\label{sx4}
\frac{d\sigma_{x}}{dt}&=& - \frac{2s_{0}}{\hbar}{F(t)}\sigma_{y}(t)
-\frac{2V_{0}}{\hbar}\cos(\omega t)\sigma_{y}(t) \nonumber\\[0.5cm]
&& +\frac{\alpha}{2}\left\{{\frac{d\sigma_{z}(t)}{dt}}\sigma_{y}(t)
+\sigma_{y}(t){\frac{d\sigma_{z}(t)}{dt}}\right\},
\end{eqnarray}
\begin{eqnarray}
\label{sy4}
\frac{d\sigma_{y}}{dt}&=&2\Delta_{0}\sigma_{z}(t) +
\frac{2s_{0}}{\hbar} {F(t)}\sigma_{x}(t)
+\frac{2V_0}{\hbar}cos(\omega t)\sigma_{x}  \nonumber\\[0.5cm]
&& -\frac{\alpha}{2}\left\{\frac{d\sigma_{z}(t)}{dt}\sigma_{x}(t)
+\sigma_{x}(t){\frac{d\sigma_{z}(t)}{dt}}\right\}, \\[1cm] 
\label{sz4} \frac{d\sigma_z}{dt}&=&-2\Delta_{0}\sigma_{y}(t), 
\end{eqnarray}
\noindent with the memory kernel $G(t-t^\prime)$=$2\alpha\delta(t-t^\prime)$.

Further simplifying equations (\ref{sx4})-(\ref{sz4}) and making use
of the properties of spin operators, we finally obtain  the Langevin
equations of motion for spin variables as
\begin{eqnarray}
\label{sx5}
\frac{d\sigma_x}{dt}&=&\left\{-\frac{2s_{0}}{\hbar}F(t)-\frac{2V_{0}}{\hbar}
\cos(\omega t)\right\}\sigma_{y}(t)
-\frac{2\alpha\Delta_{0}}{\hbar}, \\[1cm]
\label{sy5} \frac{d\sigma_y}{dt}&=&2\Delta_{0}\sigma_{z}(t)
+\left\{\frac{2s_{0}}{\hbar}F(t)+\frac{2V_{0}}{\hbar}
\cos(\omega t)\right\}\sigma_{x}(t),\\[1cm]
\label{sz5} \frac{d\sigma_z}{dt}&=&-2\Delta_{0}\sigma_{y}(t).
\end{eqnarray}

The above Langevin equations involve the operator random force
$F(t)$. The statistical properties of $F(t)$ can be obtained
using the equilibrium distribution for bath variables,
\begin{equation}
\label{boltz}
\lan{a^\dagger_{k}a_{k}}\ran=\frac{1}{e^{\beta\hbar\omega_{k}}-1},
\end{equation}
where $\beta=1/k_{B}T$. Using this and the Ohmic spectral density for the
bath oscillators, the symmetrized autocorrelation for the
operator valued random force $F(t)$ is given by[12,13],

\begin{equation}
\label{scf}
\frac{1}{2}\lan{{F(t)}{F(t^\prime)}+{F(t^\prime)}{F(t)}}\ran=
\sum g_{k}^2\cos\omega_{k}(t-t^\prime)
\coth(\hbar\omega_{k}/2k_{B}T)
\end{equation}
\noindent and the nonequal time commutator is given by
\begin{equation}
\label{neqt}
[F(t),F(t^\prime)]=-2i\sum{g_{k}^{2}\sin{\omega_{k}(t-t^\prime)}}.
\end{equation}

III. {\bf The Spin Bloch equation and their solutions.}

\par Owing to the operator nature of the random Langevin force $F(t)$,
it is difficult to solve for the expectation values of spin
variables using equations (\ref{sx5})-(\ref{sz5}).
 For simplification we make a first
approximation in that the operator random force is treated as a
classical c-number random variable. One can readily verify that in the
classical limit [12,13], taking $\hbar\rightarrow 0$, the nonequal
time commutator of $F(t)$ vanishes and the autocorrelation of
the Gaussian random force $F(t)$ becomes 
\begin{equation}
\label{scfc}
\lan{{F(t)}{F(t^{\prime})}}\ran=\eta kT\delta(t-t^{\prime}),
\end{equation}
\noindent where $\eta$ is the friction coefficient and is related to
Kondo parameter $\alpha$ through the
following realation
\begin{equation}
\label{etal}
\eta=(\hbar/2s_{0}^{2})\alpha.
\end{equation}
\par The classical Markov approximation for $F(t)$ is valid in
the high temperature limit, which will become clear later. With
the above approximation one can readily write down the equations
of motion of spin variables averaged over the ensemble
of realizations of random fluctuations $F(t)$, whose
autocorrelation is given by equation (\ref{scfc}), with the help of
Novikov's theorem [14-16]. We get

\begin{eqnarray}
\label{sx6}
\frac{d\lan{\sigma_{x}(t)}\ran}{dt}&=&-\delta\lan{\sigma_{x}(t)}\ran
-2\alpha\Delta_{0}-A\cos(\omega t)
\lan{\sigma_{y}(t)}\ran, \\[1cm]
\label{sy6} \frac{d\lan{\sigma_{y}(t)}\ran}{dt}&=&
2\Delta_{0}\lan{\sigma_{z}(t)}\ran
-\delta\lan{\sigma_{y}(t)}\ran+A\cos(\omega t)
\lan{\sigma_{x}(t)}\ran,\\[1cm]
\label{sz6} \frac{d\lan{\sigma_{z}(t)}\ran}{dt}&=&-2\Delta_{0}\lan{\sigma_{y}(t)}\ran,
\end{eqnarray}
\noindent where $\delta=2{\alpha k_{B}T}/{\hbar}$, 
and $A=2V_0/\hbar$. Equations (\ref{sx6}) to (\ref{sz6}) represent the Bloch
equations for spin variables. Unlike in the standard NMR situation, there
is no relaxation term in the evolution of ${\lan{\sigma_z}\ran}_0$.
This is because fluctuating environmental fields are exclusively along the
$z$ direction [10].
These equations are characterized by
a single relaxation time $\tau=\delta^{-1}={\hbar/2\alpha k_{B}T}$. In
the absence of an external driving field, the system
will relax asymptotically to equilibrium and the
equilibrium value of population
difference between the two levels separated by an energy value
$2\hbar\Delta_0$ is given by
$\lan{\sigma_x}\ran =(\hbar\Delta_0/k_{B}T)$
instead of $\tanh(\hbar\Delta_0/k_{B}T)$.
This shows that our approximation of treating the operator
random force as classical random c-number variable is valid for high
temperature such that ${\hbar\Delta_0/k_{B}T}\ll {1}$. The stationary
solution of Bloch equations in the presence of an external field
can be found by using the method of harmonic balance. For this we
assume stationary solutions for ${\lan{\sigma_z}\ran}_s$ and
${\lan{\sigma_x}\ran}_s$ to have the form
\begin{eqnarray}
\label{sol1}
{\lan\sigma_{z}(t)\ran}_s=a\cos(\omega t)+b\sin(\omega t) \nonumber\\[0.5cm]
\equiv {\lan\sigma_{z}(t)\ran}_{0}\cos(\omega t -\phi)
\end{eqnarray}
\begin{equation}
\label{sol2}
{\lan\sigma_{x}(t)\ran}_s=y+c\cos(\omega t)+d\sin(\omega t)
\end{equation}
\noindent where $a$,$b$,$y$, $c$ and $d$ are constants to be determined.
Substituting (\ref{sol1}) and (\ref{sol2}) in the Bloch
equations (\ref{sx6}) to (\ref{sz6}) and using harmonic balance one readily
gets an expression for the required steady state amplitude
${\lan{\sigma_z}\ran}_0$ for the average position of the
particle $\lan{\sigma_z}\ran$ as

\begin{eqnarray}
{\lan{\sigma_z}\ran}_0&=&\sqrt{a^2+b^2} \nonumber \\[0.5cm]
&& =4V_{1}\Delta_{1}^2\frac{\sqrt{(\omega_{1}^2-4\Delta_{1}^2)^2+
4\alpha^2\omega_{1}^2}}
{(\omega_{1}^2-4\Delta_{1}^2)^2+4\omega_{1}^2\alpha^2+2V_{1}^2\omega_{1}^2}.
\end{eqnarray}
\noindent Here in the above expression we have rescaled all the
energy variables in terms of $k_{B}T$, i.e.,
 $V_{1}=V_{0}/k_{B}T$, $\Delta_{1}=\hbar\Delta_{0}/k_{B}T$
and $\omega_{1}=\hbar\omega/k_{B}T$ are dimensionless variables.
Note that in the absence of driving field the amplitude
${\lan{\sigma_z}\ran}_0$  vanishes
as expected. Away from the resonance ($\omega=2\Delta_0$), the
oscillation amplitude for high frequency field scales as
($1/{\omega^2}$). In the limit of low frequency (static) driving
$\omega\rightarrow 0$ and for small $V_0$ , i.e., in the
adiabatic limit ${\lan{\sigma_z}\ran}_0$ is independent of
relaxation time $\tau$ $(={\hbar/2\alpha k_{B}T}$). These results are
consistent with  those obtained in ref. [8,9].

IV. {\bf Results and discussion}

\par In fig.(1) we have plotted the stationary amplitude of the
average position of the particle ${\lan{\sigma_z}\ran}_0$ as a
function of the dissipation coefficient (or Kondo parameter) $\alpha$,
for  various values of the dimensionless field amplitude $V_{1}$
($\equiv {V_0/k_{B}T}$). 
We have restricted to the case of resonant condition
$\omega=2\Delta_0$. The solid curve is for $V_{1}=0.01$, long dashed
curve for $V_{1}=0.04$ and small dashed curve for $V_{1}=0.07$.
We see that all these curves exhibit maxima at an optimum value of
$\alpha$, which depends on $\Delta_{1} , V_1$ and $\omega_{1}$. The
maximum value of the peak ($M_p$), however, is independent of
field strength and is given by
$M_p=\sqrt{2}(\Delta_{1}^2/\omega_{1})$, but the position of the
peak ($\alpha_{m}$=$\sqrt{2V_{1}^{2}\omega_{1}^{2}
-(\omega_{1}^{2}-4\Delta_{1}^{2})^{2}}/2\omega_{1}$)
 shifts towards higher values of $\alpha$ as we increase $V_1$.
The independence of the value of peak maxima on the field strength is also
noted in reference [8,9], for resonant case. Our expression for
$M_p$ is valid for any frequency.
The occurrence of the peak or the maximum in the
${\lan{\sigma_z}\ran}_0$ as a function of coupling strength
$\alpha$ is attributed to stochastic resonance in quantum two-level
systems. This is a result of cooperative phenomena between
various competing mechanisms of energy exchange between
the two-state system with thermal bath and the external driving field.

\par In fig.(2) and fig.(3) we have plotted
${\lan{\sigma_z}\ran}_0$ versus $\omega_{1}(\equiv{\hbar\omega/k_{B}T})$.
For this we have taken $V_{1}(\equiv{V_{0}}/k_{B}T)$=0.04 and 
$\Delta_{1}(\equiv\Delta/k_{B}T)$=0.2. For
fig.(2) $\alpha=0.05$ and for fig.(3) $\alpha=0.01$. Fig.(2) and (3) should
be compared with fig.(7a) and fig.(7b) of respectively ref.[9] for their
qualitative behaviour. In fig.(2) ${\lan{\sigma_z}\ran}_0$ exhibits a maximum
at the resonant frequency $\omega=2\Delta_{0}$, whereas two maxima
are seen in fig.(3) at off resonance frequency values. This
indicates  stochastic resonance can be obtained even for the off
resonance conditions. This further indicates that the stochastic
resonance is indeed a bonafied resonance [17,18].

\par In fig(4) we have plotted ${\lan{\sigma_z}\ran}_0$ as a
function of external driving field amplitude
$V_{1}(\equiv{V_{0}/k_{B}T})$ for various values of the dissipation
parameter, the frequency of the field is $\omega_{1}=2\Delta_{1}=0.4$.
The solid, long dashed and small dashed curves are for
$\alpha=(0.02),(0.04)$ and (0.06),
respectively. We notice that the response of the system, i.e.,
${\lan{\sigma_z}\ran}_0$, initially increases with the field
amplitude and attains a maximum value at a particular value of
$V_1$ depending on $\alpha$ and on other parameters. 
After exhibiting the maximum
${\lan{\sigma_z}\ran}_0$  decreases as we increase $V_1$
further. The maximum value ($M_p$) of the peak in the  ${\lan{\sigma_z}\ran}_0$
for a given value of $\Delta_1$, $\omega_{1}$ and $\alpha$ occurs at the
field amplitude value
$V{_1}=\sqrt\frac{\{(\omega_{1}^2-4\Delta_{1}^2)^2+4\alpha^{2}\omega_{1}^2\}}
{2\omega_{1}^2}$ and is equal to
$M_p=
\sqrt{2}(\frac{\Delta_{1}^2}{\omega_{1}})$
which is independent of $\alpha$, the Kondo parameter. 
The linear response theory is valid for small $V_{1}$ to
the left side  regime of the maximum. In these plots stochastic
resonance manifests as a breakdown of linear response theory,
thus bringing out the non linear nature of the problem
explicitly. 

In fig.(5) we have plotted ${\lan\sigma_{z}\ran}_0$ versus
$\Delta_{1} (\equiv \hbar\Delta_0/k_{B}T)$, for given
$\omega_{1}(\equiv \hbar\omega/k_{B}T)$=0.4 and a small $\alpha$=0.04.
The maximum appears well within the range of acceptable values of
$\Delta_{1} \ll {1}$, i.e., $\hbar\Delta_{0} \ll k_{B}T$.
For very small values of $\alpha$ we do get maximum response at
two values of $\Delta_1(\neq 2\omega_1)$ indicating stochastic
resonance at off resonance condition (similar to the observation
made in fig. (3)).

In conclusion we have derived quantum Langevin equation for a
dissipative two-level system, driven by monochromatic ac field,
starting from microscopic spin-boson Hamiltonian. The equations
of motion for the average values of the spin variables are obtained
in the high temperature limit. We have obtained an analytical
expression for the amplitude of coherent oscillation in the
average position of the particle, which exhibits stochastic
resonance with respect to various parameters in the problem.
These results are in qualitative agreement with the results obtained
by Makarov and Makri using the numerical method of iterative
path integral scheme [8,9].

%\vfill
%\eject
\newpage
{\bf Figure caption}

Fig.1 Plot of the stationary amplitude of average position
${\lan{\sigma_z}\ran}_{0}$ versus disipation coefficient (Kondo
parameter) $\alpha$ for a fixed value of $\omega$=$2\Delta_0$=0.4.
For this case $V_{1}$($\equiv{V_0/k_{B}T}$)=0.01 (solid curve),
0.04(long dashed curve) and 0.07(small dashed curve).
\vspace{0.1in}

Fig2. Plot of the stationary amplitude of average position
${\lan{\sigma_z}\ran}_{0}$ versus $\omega_{1}$
($\equiv{\hbar\omega_{0}/k_{B}T}$) for $V_1$=0.04,
$\Delta_{1}$=0.2 and $\alpha$=0.05.

\vspace{0.1in}
Fig3.  Plot of the stationary amplitude of average position
${\lan{\sigma_z}\ran}_{0}$ versus $\omega_{1}$
($\equiv{\hbar\omega_{0}/k_{B}T}$) for $V_1$=0.04,
$\Delta_{1}$=0.2 and $\alpha$=0.01.
\vspace{0.1in}

Fig.4 Plot of the stationary amplitude of average position
${\lan{\sigma_z}\ran}_{0}$ versus  $V_{1}$($\equiv{V_0/k_{B}T}$)
for a fixed value of $\omega$=$2\Delta_0$=0.4.
For this case $\alpha$=0.02 (solid curve),
0.04(long dashed curve) and 0.06(small dashed curve).
\vspace{0.1in}

Fig.5 Plot of stationary amplitude of average position
${\lan{\sigma_z}\ran}_{0}$ versus
$\Delta_1$ $\equiv$ ($\hbar\Delta_0/k_{B}T$) for a fixed value of
$V_1$=0.04, $\omega_1$ =0.4, and $\alpha$ =0.04.
\end{document}